# On-Chip Silicon Waveguide Bragg Grating Photonic Temperature Sensor

Nikolai N. Klimov[1,3], Sunil Mittal[2,3,4], Michaela Berger[1], Zeeshan Ahmed[1]

[1]Thermodynamic Metrology Group, Sensor Science Division, Physical Measurement Laboratory, National Institute of Standards and Technology, Gaithersburg, MD, USA

[2]Quantum Optics Group, Quantum Measurement Division, Physical Measurement Laboratory, National Institute of Standards and Technology, Gaithersburg, MD, USA

[3]Joint Quantum Institute, University of Maryland, College Park, MD, USA

[4]Department of Electrical and Computer Engineering, University of Maryland, College Park, MD, USA

*Corresponding author: zeeshan.ahmed@nist.gov

Abstract: Resistance thermometry is a time-tested method for taking temperature measurements. In recent years fundamental limits to resistance-based approaches spurred considerable interest in developing photonic temperature sensors as a viable alternative. In this study we demonstrate that our photonic thermometer, which consists of a silicon waveguide integrated with a Bragg grating, can be used to measure temperature changes over the range from 5 °C to 160 °C with a combined expanded uncertainty [k = 2 ; 95% confidence level] of 1.25 °C. The computational modeling of the sensor predicts the resonance wavelength and effective refractive index within 4% of the measured value.

Though today's resistance thermometers can routinely measure temperatures with uncertainties of ≤ 10 mK,[1] they are sensitive to environmental variables such mechanical shock and humidity, which cause the sensor resistance to drift over time requiring expensive, time consuming calibrations. The desire to reduce sensor ownership cost has produced considerable interest in the development of photonic temperature.[2–4] Photonic temperature sensors have the potential to leverage advances in frequency metrology to provide cost effective measurement solutions. Proposed sensor technologies range from temperature sensitive dyes[5] to macro scale fiber Bragg grating (FBG).[2,3,6,7] It has been shown that FBGs have a temperature dependent shift of ≈ 0.01 nm/°C.[2,3,6,7] The uncertainty for the best fit for temperature-wavelength relationship using FBG sensors has been reported to vary from 0.55 °C to 2 °C (1σ).[8] We recently demonstrated that micron scale silicon ring resonators, when interrogated using *swept wavelength* methodology, can resolve temperature changes of 1 mK, while in the *side of fringe, constant power mode*, the noise floor is further lowered to 80 μK.[9–15]

In this study we examine the temperature dependent response of our nanoscale silicon waveguide Bragg grating (Si WBG) photonic sensor over the range of 5 °C to 160 °C. Our results demonstrate that Si WBG thermometer's temperature response (Δλ/ΔT) is ≈8x better than FBG based sensor.

Figure 1a shows the design of the Si WBG photonic sensor with its relevant dimensions. This sensor consists of a Si straight-probe waveguide with cross-section of 510 nm × 220 nm. A 330 μm-long active region of the waveguide has Bragg grating in the square-wave form modulation of the waveguide width. The silicon waveguide fundamental transverse mode $\left(E_y^2 + E_z^2\right)^{1/2}$, which we calculated using finite difference eigenmode (FDE) solver in Lumerical[1] software, is shown on Fig. 1b. The silicon photonic sensors were made at CEA-LETI (Laboratoire d'Electronique et de Technologie de l'Information, France) fabrication facility using standard CMOS compatible manufacturing technology. The Bragg waveguide is probed using a custom built interrogation system[15] (Fig. 1c). In this setup a C-band laser (New Focus TLB-6700 series) is swept over the sensor resonance. Ten percent of laser power was immediately picked up from the laser output for wavelength monitoring (HighFinesse WS/7) while the rest, after passing through the photonic device, was detected by a large sensing-area power meter (Newport, model 1936-R). Light was coupled into and out of the waveguide using grating couplers. Unless noted otherwise, all experiments where performed with 1 mW of incident laser power.

---

[1] Disclaimer: Certain commercial fabrication facility, equipment, materials or computational software are identified in this paper in order to specify device fabrication, the experimental procedure and data analysis adequately. Such identification is not intended to imply endorsement by the National Institute of Standards and Technology, nor is it intended to imply that the facility, equipment, material or software identified are necessarily the best available.

Figure 1d shows resonance transmission spectra in Si WBG sensors measured at different temperatures. At the lowest measured temperature of 5 °C the Bragg wavelength, $\lambda_B$, is centered at 1534.24 nm with a resonance bandwidth $\Delta\lambda \approx 5.13$ nm and isolation depth of $\approx 8$ dB. The large bandwidth is attributed to strong refractive index modulation that is intrinsic to the device design used here. We observed that over the temperature range from 5 °C to 160 °C, the Bragg wavelength shows a monotonic upshift with temperature. At temperatures below 80 °C, $\lambda_B(T)$ can be approximated with a linear temperature dependence with a slope of $\partial\lambda_B/\partial T \approx 0.082$ nm/°C (Fig. 1e).

At temperatures above 80 °C, the Bragg wavelength temperature dependence deviates from linearity and shows a weak quadratic dependence on temperature (for more information see supplementary material, Fig. S1 and S2). On the other hand, the peak bandwidth changes only slightly increasing as temperature is increased at a slow rate of the order of $\partial(\Delta\lambda)/\partial T \approx 0.00086$ nm/°C (Fig. 1f). In the strong-grating limit, the slow change in bandwidth indicates the index modulation ($\delta n/n_{eff}$) is slowly varying function of temperature. The slow variation in index modulation is likely due to small differences in the temperature dependence of silica and silicon's thermo-optic coefficient. The effective refractive index is predominantly determined by silicon, while the $\delta n$ is difference in refractive between silicon and silica that make up the Bragg grating. We also confirm that increasing incident laser power from 1 mW to 3.6 mW impacts neither the Bragg wavelength nor the resonance bandwidth (Supplementary material, Fig. S3), indicating that a self-heating is negligible in laser power range employed here. These results clearly indicate the observed thermal response of the Bragg waveguide derives from the first order molecular susceptibility tensor.[16] Higher order effects such as two-photon absorption which contributes to self-heating[17] can be ignored. A practical implication of this result is that calibration of Si WBG can be treated as being independent off laser power.

We also examine the system hysteresis upon thermo-cycling (Figs. 1e and 1f). Solid red circles and open blue squares on $\lambda_B(T)$ (Fig. 1e) and $\Delta\lambda(T)$ (Fig. 1f) data sets correspond to temperature dependences as $T$ is increased (from 5 °C up to 160 °C) and then decreased (from 160 °C down to 5 °C), respectively. Thermo-cycling measurements of both $\lambda_B(T)$ and $\Delta\lambda(T)$ show no thermal hysteresis within the experimental uncertainty. We estimate measurement uncertainties for $\lambda_B$ and $\Delta\lambda$ from Figs. 1g and 1h, where for every measured temperature we plotted the deviations of $\lambda_B$ and $\Delta\lambda$ from their average values, $\lambda_{B,ave} = 0.5(\lambda_{B,heating} + \lambda_{B,cooling})$ and $\Delta\lambda_{ave} = 0.5(\delta\lambda_{heating} + \delta\lambda_{cooling})$. The corresponding uncertainties for $\lambda_B$ and $\Delta\lambda$ are $\approx 0.01$ nm and $\approx 0.025$ nm, respectively. Based on Bragg wavelength sensitivity to temperature variation ($\partial\lambda_B/\partial T \approx 0.082$ nm/°C) we estimate best fit uncertainty is $\approx 0.1$ °C. The combined expanded uncertainty of the present measurement is 1.25 °C, which is dominated by uncertainty in peak center measurement and temperature measurement (Table 1). Development of narrower bandwidth devices that allow more accurate peak center measurement will be a focus going forward.

The Bragg grating is characterized by two main parameters: resonance wavelength ($\lambda_B$) and bandwidth ($\Delta\lambda$). For a Bragg grating (Fig. 1a) with a grating period ($\Lambda$), and refractive indecies of valleys ($n_a$) and peaks ($n_b$) components of the Bragg grating the Bragg wavelength of the grating is given by:

$$\lambda_B = 2\Lambda n_{eff} \qquad (1)$$

where, $n_{eff} = a\, n_a + b\, n_b$, is effective refractive index of the structure, while $a$ and $b$ denotes the grating duty cycle (for current Si WBG $a = b = 0.5$). The bandwidth of a Bragg waveguide is given by:

$$\frac{\Delta\lambda}{\lambda} = \sqrt{\left(\frac{\upsilon\overline{\delta n_{eff}}}{n_{eff}}\right)^2 + \left(\frac{2}{N}\right)^2} \qquad (2)$$

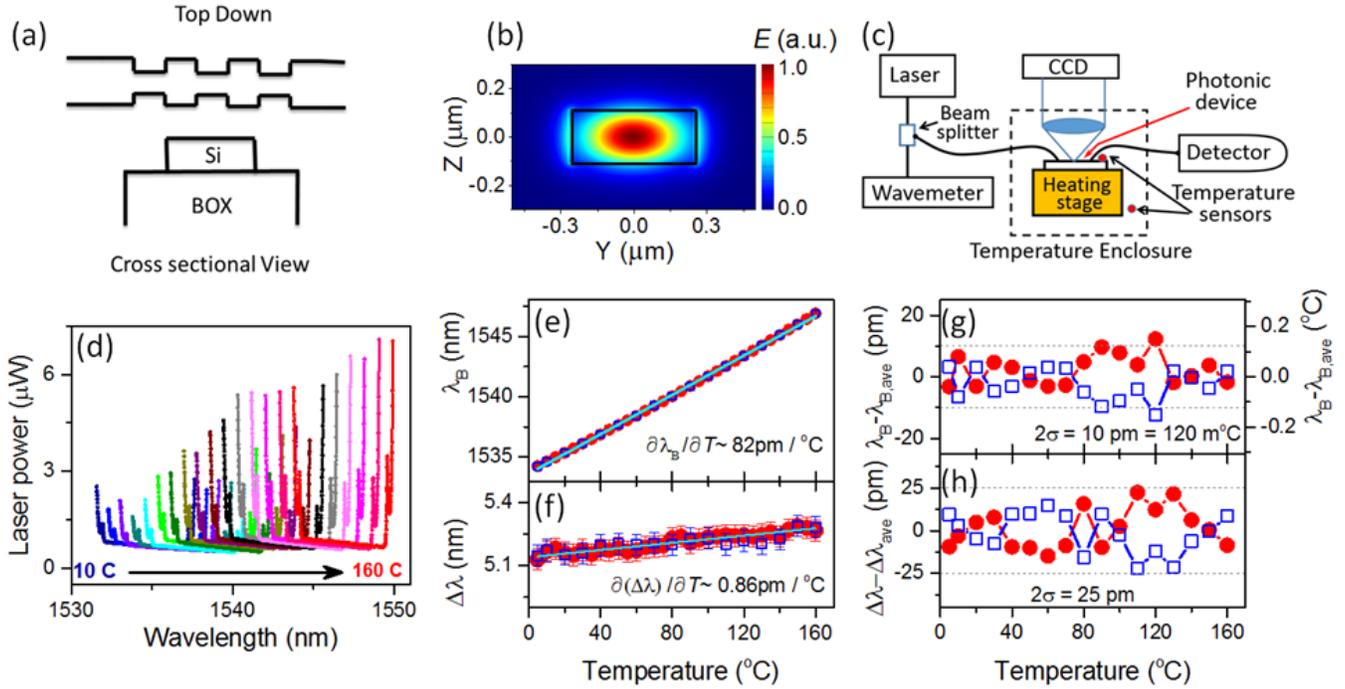

Figure 1: Color online. (a) Schematics of Bragg grating in Si WBG sensor. The device parameters are $H = 220$ nm, $W_a = 460$ nm, $W_b = 510$ nm, $L_a = L_b = 165$ nm, $N_{periods} = 1000$, buried oxide (BOX) and plasma-enhanced chemical vapor deposited (PECVD) SiO$_2$ thicknesses are 2 µm and 1.6 µm, respectively. (b) finite difference eigenmode solver (FDE) calculated transverse mode in a Si waveguide with 220 nm × 510 nm cross-section. (c) A block diagram of the microscopy-based interrogated setup to interrogate the photonic devices. (d) Si WBG resonance transmission spectra measured at different temperatures between 5 °C and 160 °C. Temperature dependence of (e) Bragg wavelength, $\lambda_B$, and (f) resonance bandwidth, $\Delta\lambda$, measured during heating (solid red circles) and cooling (open blue squares). Straight lines in (e, f) are linear fits with slopes of ≈ 0.082 nm/°C (e) and ≈ 0.00086 nm/°C (f). Measurement uncertainty in (e) and (f) is 0.05 nm. Temperature dependence of the deviation of (g) $\lambda_B$ and (h) $\Delta\lambda$ from their average values, $\lambda_{B,ave} = 0.5(\lambda_{B,heating} + \lambda_{B,cooling})$ and $\Delta\lambda_{ave} = 0.5(\Delta\lambda_{heating} + \Delta\lambda_{cooling})$, measured during heating (solid red circles) and cooling (open blue squares). Measurement uncertainty in (g, h) is 0.05 nm.

where $\upsilon\overline{\delta n_{e_{ff}}}$ is the "AC" component of the refractive index modulation, $N$ is number of periods, and $\Delta\lambda/\lambda$ is the normalized bandwidth. In our FDTD modeling the calculated resonance Bragg wavelength, effective refractive index and its modulation are $\lambda_B^{calc} \approx 1597.20$ nm, $n_{eff}^{calc} \approx 2.42$ and $\Delta n \approx 0.09$, respectively. Given a Bragg grating period of $L = 330$ nm and experimentally observed Bragg wavelength of $\lambda_B(20\,°C) = 1535.35$ nm, the corresponding measured effective index of Si WBG calculated from Eq. 1 is $n_{eff}^{exp} \approx 2.33$, which is within 4% of the theoretically calculated value $n_{eff}^{calc}$. The difference in theoretical and experimentally calculated refractive index and by extension bandwidth and resonance wavelength derives in part from fabrication errors[18]. The broad measured bandwidth ($\Delta\lambda(20\,°C) \approx 5.16$ nm) indicates the ridge structure creates a strong refractive index modulation. The bandwidth of the Bragg reflection can be reduced by reducing the sidewall corrugation depth, altering the ridge profile, and/or employing more complex waveguide geometry such as slab on a ridge geometry.[19] Similar devices and experiments have been also demonstrated in Refs.[20,18].

In summary, we have demonstrated that Si WBG sensor fabricated using CMOS compatible manufacturing technology is a viable photonic temperature sensing solution. Our results demonstrate these devices can be used to measure temperature at least over the range from 5 °C to 160 °C with a temperature sensitivity of ≈ 0.082 nm/°C, a factor of ≈ 8x improvement over FBG based sensor. In contrast to a ring resonator sensor that can readily track temperature variation only over a relatively narrow interval ($\Delta T \approx 100$-150 °C) limited by the free spectral range of the device (distance between adjacent resonance peaks), Si WBG sensor with its well-isolated resonance peak can continuously measure $T$ from cryogenic (≈ -269 °C) up to silica melting point (≈ 1600 °C) temperatures. The current uncertainty of temperature measurements (1.25 °C) using our photonic Si WBG sensors can be further improved by fabrication of narrower bandwidth devices which will reduce the uncertainty in peak center determination, a major component of the uncertainty budget (Table 1).

Table 1:
Combined Uncertainty of Si WBG Temperature Sensor

| Component | Uncertainty (°C) |
|---|---|
| Temperature | 0.075 |
| Peak Center | 0.61 |
| Best Fit | 0.1 |
| Combined Uncertainty | 0.62 |
| Expanded Uncertainty ($k=2$) | 1.25 |

Appendix

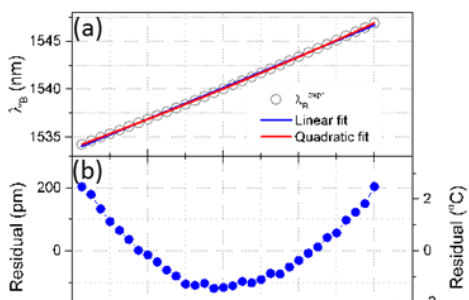

Figure S 1: (a) Temperature dependence of Bragg wavelength, $\lambda_B$ measured during heating. Measurement uncertainty ($k=1$) is 0.05 nm. Blue and red curves are linear and quadratic fits to the experimental data. (b) Residual to linear fit in (a). (c) Residual to quadratic fit in (a).

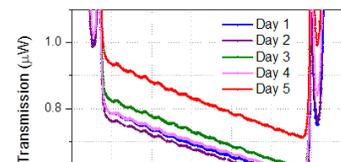

Figure S 2: Reproducibility of the measurement (k=1). (a) Si WBG resonance transmission spectra measured at a fix temperature of 20 °C, but on different days. (b) Reproducibility of Bragg wavelength, $\lambda_B$, measured on different days. (c) Reproducibility of resonance bandwidth, $\Delta\lambda$, measured on different days. Red circles in (b) and (c) represent absolute values of $\lambda_B$ and $\Delta\lambda$ in nm, while green squares correspond to relative change (in pm) of $\lambda_B$ and $\Delta\lambda$ from their average values $\lambda_{B,ave}$ and $\Delta\lambda_{ave}$.